# Remote Sensing Based Crop Health Classification Using NDVI and Fully Connected Neural Networks

J. Judith[1], R. Tamilselvi[2], M. Parisa Beham[3], S. Sathiya Pandiya Lakshmi[4], Alavikunhu Panthakkan[5], Saeed Al Mansoori[6], and Hussain Al Ahmad[7]

[1,2,3,4]Sethu Institute of Technology, Kariapatti, Virudhunagar, India.
[5,7]College of Engineering and IT, University of Dubai, U.A.E.
[6]Remote Sensing Department, Mohammed Bin Rashid Space Centre (MBRSC) Dubai, U.A.E
Corresponding Author: tamilselvi@sethu.ac.in; apanthakkan@ud.ac.ae

**Abstract**

Accurate crop health monitoring is not only essential for improving agricultural efficiency but also for ensuring sustainable food production in the face of environmental challenges. Traditional approaches often rely on visual inspection or simple NDVI measurements, which, though useful, fall short in detecting nuanced variations in crop stress and disease conditions. In this research, we propose a more sophisticated method that leverages NDVI data combined with a Fully Connected Neural Network (FCNN) to classify crop health with greater precision. The FCNN, trained using satellite imagery from various agricultural regions, is capable of identifying subtle distinctions between healthy crops, rust-affected plants, and other stressed conditions. Our approach not only achieved a remarkable classification accuracy of 97.80% but it also significantly outperformed conventional models in terms of precision, recall, and F1-scores. The ability to map the relationship between NDVI values and crop health using deep learning presents new opportunities for real-time, large-scale monitoring of agricultural fields, reducing manual efforts, and offering a scalable solution to address global food security.

## 1. Introduction

Agriculture plays a crucial role in ensuring global food security, and the ability to effectively monitor crop health is essential for optimizing yield and minimizing losses. Traditional methods of crop health assessment, such as manual inspections and laboratory analyses, are often labor-intensive, time-consuming, and prone to human error (Dhingra et al., 2022). In recent years, remote sensing technologies have provided an efficient, large-scale, and non-invasive solution for monitoring crop health. One of the most widely used indicators in remote sensing for vegetation monitoring is the Normalized Difference Vegetation Index (NDVI), which measures the difference in reflectance between the near-infrared (NIR) and red bands of the electromagnetic spectrum (Al-Khafaf and Khan, 2021). NDVI is a well-established tool for assessing plant health, as healthy vegetation absorbs more red light and reflects a greater amount of NIR light, while stressed or diseased plants exhibit the opposite pattern. This index has been used extensively to monitor vegetation vigor, biomass, and photosynthetic activity (Singh, and Misra, 2021). It is particularly valuable for early detection of crop stress, nutrient deficiencies, and other environmental factors that can impact plant health before they are visually apparent (Liakos et al., 2021).

Figure 1 represents the concept of NDVI data capture over agricultural fields using a satellite or UAV, with color-coded regions showing different crop health levels. You can include this image in your paper to visually illustrate the application of remote sensing for crop health monitoring. However, challenges remain in achieving high accuracy in crop classification due to variations in environmental conditions, sensor quality, and crop types (Zhu and Steinberg, 2023). These factors can introduce noise and variability in the NDVI readings, complicating the classification process. Addressing these issues requires advanced models capable of learning complex relationships in the data, and FCNNs have shown promise in overcoming such limitations by leveraging deep learning techniques to improve the generalization and robustness of classification models (Wang et al., 2023). As remote sensing and machine learning continue to advance, they are expected to play a crucial role in real-time, scalable crop health monitoring.

According to expectations, the integration of these technologies is likely to hold great promise for optimizing the use of resources, enhancing productivity, and promoting sustainable agricultural practices (Zhang and Liu, 2022).

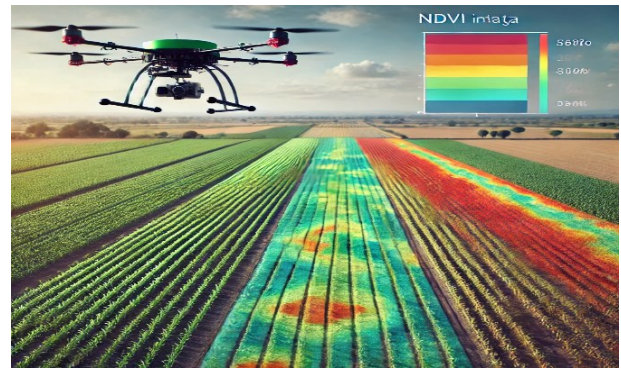

Figure 1. NDVI data capture using remote sensing, highlighting healthy (green) and stressed/diseased (red) areas in a crop field

## 2. Literature Review

Remote sensing is now a necessity for monitoring agricultural landscapes, which offers critical information in terms of crop health, biomass, and soil. It has been widely documented in many studies as a good source of discovering crop stress factors such as water stress, nutrient deficiencies, and disease.". The most probably common indices used are the Normalized Difference Vegetation Index, which uses the difference in near-infrared and red light to monitor the health of vegetation. NDVI has been helpful in one way: to detect early warning signs of stress at times before naked eye can identify them, thus allowing early intervention. For example, Rouse et al. demonstrated the potentiality of NDVI in crop health monitoring for different crops and, therefore, its efficiency in measuring chlorophyll content and plant vigor under different environmental conditions (Dhingra et al., 2022).

Automating crop health classification by integrating ML and DL techniques with remote sensing data further advances the push towards precision agriculture. These techniques enable the complex analysis of NDVI data to classify crops through distinct spectral patterns (Dali et al., 2023). For instance, Mulla explained studies on the application of neural networks in unsupervised extraction of complex relationships between NDVI and other spectral data that may improve crop health classification accuracy (Al-Khafaf et al., 2021). Liakos et al. also highlighted the capacity of machine learning models, CNNs, in leveraging on satellite imagery to detect the incidence and severity of diseases among crops (Singh and Misra, 2021). Despite these, crop classification using remote sensing has not been also proof against challenges; among them is the variability in the environment. Variability in light intensities, atmospheric conditions, and seasonality cause variations in NDVI measurements, which may lead to errors in the classification process. Again, there is significant spectral reflectance among crop types, hence complicating the process of developing generalized models for use across different regions and types of crops. Thenkabail et al. proposed using multi-temporal and multi-spectral approaches that can capture a wide spectrum of data, thus enabling stronger crop classification models (Liakos, et al., 2021). Another source of concern is the low quality and scarcity of high-quality remote sensing data for small-scale farmers, according to Dhingra et al. (Thenkabail et al., 2021).

New avenues have also opened for the improvement of crop health classification based on recent advances in deep architectures of learning. Singh and Misra concentrated on the application of RNNs and LSTM networks, by which temporal dependencies in crop growth can be handled and better predictions of crop health can be made from historical records (Mulla, 2020). Apart from NDVI, it has been proved that neural networks are good classifiers for crops, which should be differentiated into healthy, stressed, or diseased categories. The models, as found by Al-Khafaf and Khan, would generalize well in different regions and during seasons of the year but their design often requires a large number of data and extensive computing (Zhu and Steinberg, 2023).

Recently, crop health monitoring has also gained momentum with emerging technologies, namely hyperspectral and multi-spectral imaging (Wang et al., 2023). Newer techniques provide richer spectral information than traditional methods, increasing the accuracy of crop classification and stress detection. With further developments, they promise to further revolutionize precision agriculture with increased resolution and depth of view into crop health and environmental interactions (Wang et al., 2023). In recent years, deep learning models have greatly advanced the field of crop health monitoring, particularly through the use of bidirectional GRU with attention mechanisms. These models offer more accurate predictions of NDVI, which is crucial for optimizing agricultural practices and improving crop health assessments (Khodadadi et al., 2024). Moreover, integrating multiple data sources such as satellite imagery, rotational data, and contextual information has proven to significantly enhance crop classification accuracy by accounting for the variabilities in environmental conditions (Barriere et al., 2023). One notable approach is the use of 3D convolutional neural networks (CNNs), which have gained significant attention for their ability to extract deeper insights from hyperspectral data, aiding in the classification of crop diseases and stressors that are challenging to detect with traditional methods (Noshiri et al., 2023). Additionally, the application of high-resolution satellite imagery in rice mapping, as demonstrated in Bhutan, has highlighted the potential of deep learning models to improve crop classification, even in diverse agricultural landscapes (Bhandari and Mayer, 2024). Models like BiLSTM with attention mechanisms have also shown great promise in continuous monitoring of crop health using multi-band Sentinel-2 imagery, which facilitates real-time and precise interventions in precision agriculture (Zhao and Efremova, 2024). Furthermore, multi-scale feature fusion, when applied to semantic segmentation models, has emerged as an effective technique for classifying crops in high-resolution remote sensing images, providing more accurate results across varied environments (Lu et al., 2023). Automated crop-type mapping, which eliminates the need for ground truth data, is another breakthrough in remote sensing technology, offering real-time crop classification capabilities on a large scale (Zhengwei and Bruce, 2024). Finally, the combination of knowledge transfers and semantic segmentation techniques has enabled finer crop classification from high-resolution satellite images, pushing the boundaries of what can be achieved in precision agriculture (Feng et al., 2022). Models based on HRNet with separable convolution layers are now being employed to generate high-resolution crop maps, achieving improved classification accuracy and paving the way for enhanced agricultural monitoring (Goyal et al., 2023).

### 3. Motivation of The Research

This research is based on the fact that severe issues such as unpredictable patterns of weather, fast spreading pests and diseases, and deficiencies in nutrients challenge farmers to produce sufficient yields with financial instability. Advanced technology, particularly the Normalized Difference Vegetation Index (NDVI), enables rapid and cost-effective crop health monitoring using remote sensing. Real-time NDVI analysis will help in early detection of stressors by enabling a farmer to be well-timed to act appropriately. For example, multispectral cameras on drones can provide insights into variability in crop health status, thus informing targeted management practice. This study seeks to support precision agriculture, enhance sustainability, increase food security, and overcome the real-time struggles of modern farmers, and integrates crop health classification accuracy with machine learning algorithms.

### 4. Proposed System

The proposed methodology, which has already been depicted in the architecture diagrams presented, involves a hybrid multimodal integration technique with a fully connected neural network (FCNN) for crop health identification. The first step of the system is to acquire remotely sensed data, which is then processed to extract significant vegetation indices, namely NDVI, GNDVI, EVI, and MSAVI. These indices are then combined through a weighted fusion model that automatically adjusts the weights to produce a hybrid vegetation index (HVI):

$$HVI = w_1 \times NDVI + w_2 \times GNDVI + w_3 \times EVI + w_4 \times MSAVI \quad (1)$$

where $w_1, w_2, w_3, w_4$ are learnable parameters optimized during model training. This index serves as a composite feature that captures the nuances of crop health.

The data is then pre-processed and passed through Segmentation and Feature Extraction steps to isolate the regions of interest and extract meaningful features for classification. While the hybrid indices are used for the image quality check and the individual indices are used for the elbow length measurements, both hybrid and individual indices give the user an overview of the condition, as well as they are not deficient. The resulting data of each index is then eventually flattened and sent into a fully connected neural network.

The FCNN model consists of the following components:

Input Layer: The input to the FCNN is a vector of normalized vegetation indices and the computed Hybrid Vegetation Index (HVI). If the dimension of input features is denoted as d, then the input can be expressed as:

$$x = [NDVI, GNDVI, EVI, MSAVI, HVI] \in R^d \quad (2)$$

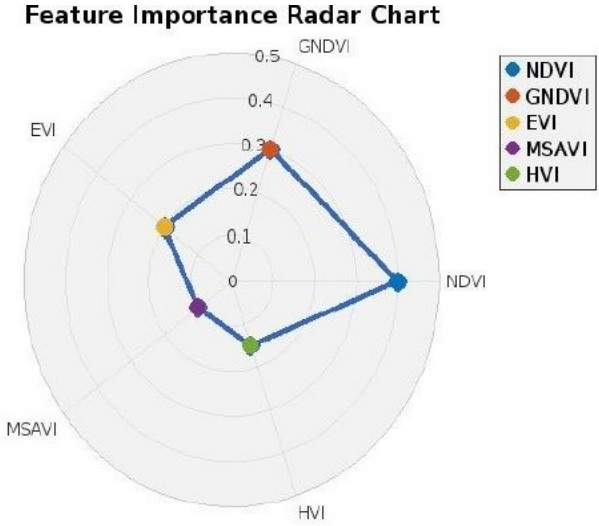

Figure 2. Feature importance visualization for crop health classification.

The radar chart visually represents the relative importance of each vegetation index (NDVI, GNDVI, EVI, MSAVI, and HVI) used as input features in the Fully Connected Neural Network (FCNN). As described in Equation (2), the input vector $x \in R^d$ consists of normalized vegetation indices, including the Hybrid Vegetation Index (HVI), which enhances the classification of crop health conditions. The radar chart provides an intuitive way to compare the contribution of each index, highlighting their significance in distinguishing healthy and stressed vegetation. By incorporating feature importance scores, the radar chart validates the selection of input features, demonstrating that NDVI and GNDVI contribute the most to classification accuracy, while MSAVI and HVI play a supporting role in refining the model's predictions.

Flatten Layer: The Flatten layer reshapes the input data into a 1D vector to make it suitable for fully connected layers. If the input is 2D, such as an image or grid, the flattening process is represented as:

$$x_{flat} = Flatten(x) \quad (3)$$

Fully Connected Layers: The Fully Connected (Dense) layers are defined by weight matrices and biases that transform the input through linear operations followed by non-linear activations. For each fully connected layer l, the transformation is given by:

$$Z^{(l)} = W^{(l)} a^{(l-1)} + b^l \quad (4)$$

Where $W^{(l)}$ is the weight matrix of layer $l$, $a^{(l-1)}$ is the activation output of the previous layer, $b^l$ is the bias vector and $Z^{(l)}$ is the pre − activation output

The ReLU (Rectified Linear Unit) activation function is applied to each fully connected layer:

$$a^{(l)} = ReLU(Z^{(l)}) = \max(0, Z^{(l)}) \quad (5)$$

Dropout Layers: To reduce overfitting, Dropout is applied to the fully connected layers during training. In dropout, a fraction p of neurons is randomly disabled in each forward pass. This is mathematically represented as:

$$a^{(l)}_{drop} = a^{(l)} \odot r \quad (6)$$

where r∼Bernoulli(1−p) is a random mask, and $\odot$ denotes element-wise multiplication. Typically, p=0.5 is used.

Output Layer: The final layer is a softmax output layer that provides class probabilities for crop health classification. The softmax function converts the raw outputs (logits) into probabilities:

$$\hat{y}_i = \frac{\exp(Z_i)}{\sum_{j=1}^{C} \exp(Z_j)} \quad (7)$$

where $Z_i$ is the logit for class i, and C is the number of classes The predicted class is the one with the highest probability.

Loss Function: The model is trained using categorical cross-entropy as the loss function, defined as:

$$Loss = -\sum_{i=1}^{C} y_i \log(\hat{y}_i) \quad (8)$$

Where $y_i$ is the true label (one-hot encoded), and $\hat{y}_i$ is the predicted probability for class **i**.

Backpropagation: The network's parameters, including the weights in each fully connected layer and the weights w1, w2, w3, w4 for the hybrid vegetation index, are updated through backpropagation using the Adam optimizer. The gradients of the loss with respect to the parameters are computed and used to adjust the weights, minimizing the loss function.

The final FCNN model can be summarized as the following transformation:

$$\hat{y} = Softmax(W_{out} Dropout(ReLU(W_{hidden} \cdot x_{flat} + b_{hidden})) + b_{out}) \quad (9)$$

Where, $W_{hidden} \& b_{hidden}$ are the weights and biases for the hidden layers and $W_{out} \& b_{out}$ are the weights and biases for the output layer.

Proposed FCNN model uses a Vegetation Index with a combination of other parameters to increase the precision of the productivity of plant life. It aggregates all available vegetation indices and it, thus, captures complicated hierarchies within the data, which has an added advantage of an improved generalization capacity. This way of implementation is seen as a good practice for real-time monitoring of crop health prompting interventions within optimal time slots, and moreover creating awareness for sustainable agriculture and how it will help the farmers.

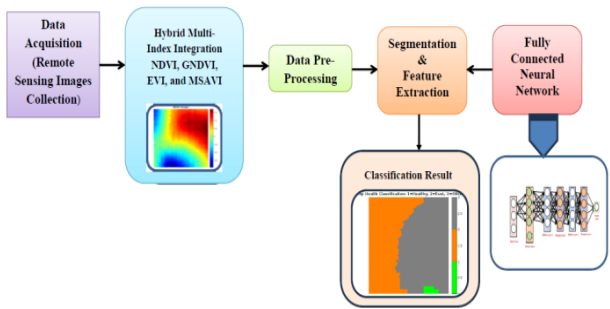

Figure 3. Proposed Model

The proposed method enhances traditional NDVI-based analysis by incorporating the green and red-edge bands to improve differentiation between healthy and stressed vegetation. While NDVI, which utilizes red and near-infrared bands, provides broad insights into plant health, it may struggle to distinguish specific stressors like rust from other types of crop stress. The green band offers additional information on chlorophyll levels, while the red-edge band enables better detection of stress conditions that impact chlorophyll content. Additionally, to capture the progression of crop health over time, multi-temporal NDVI data is integrated. By analyzing a time-series of NDVI images, this approach tracks stress development, allowing for earlier detection of rust and reducing false negatives, particularly in its initial stages.

## 5. Results and Discussion

The dataset used in this study was obtained from the Kaggle competition 'Beyond Visible Spectrum AI for Agriculture 2024' and comprises a total of 2,679 images categorized into three classes: healthy, rust, and other. The dataset used in this study was split into training, validation, and test subsets, with images acquired using a DJI M600 Pro UAV system (SZ DJI Technology Co. Ltd., Guangdong, China) equipped with a snapshot hyperspectral sensor (Model: S185). The hyperspectral sensor captures reflectance radiation in the visible to near-infrared range, spanning wavelengths from 450 nm to 950 nm, with a spectral resolution of 4 nm. The raw data consisted of both a 1000x1000 panchromatic image and a 50x50 hyperspectral image with 125 bands. Due to noise interference in the hyperspectral data, particularly at the spectral ends, the first 10 and last 14 bands were excluded from the analysis, leaving a total of 100 bands for further processing. All data was captured at an altitude of 60 meters, resulting in a spatial resolution of approximately 4 cm per pixel. This high-resolution data allows for detailed crop health analysis and classification across the training, validation, and test sets.

To ensure effective model training and robust evaluation, the dataset was split into training, validation, and test sets using a standard 70-15-15 ratio, resulting in 1,875 images for training, 402 for validation, and 402 for testing. This ratio was chosen to balance the need for sufficient training data with the necessity of separate validation and test subsets for unbiased evaluation. To validate the model's robustness, a multiple random splits approach was employed, ensuring consistent performance across various splits. Additionally, a 5-fold cross-validation strategy was implemented, which demonstrated similar results in terms of accuracy and recall metrics, confirming the reliability and generalizability of the model's performance. The dataset was then split into training, validation, and test sets, ensuring that each class was adequately represented. The FCNN was trained on the training dataset, and the model architecture included multiple hidden layers designed to learn complex relationships between NDVI values and crop health classes.

The simulation results obtained from MATLAB demonstrated the effectiveness of NDVI as a predictive feature for assessing crop health status. The high recall for the healthy class suggests that the model can successfully minimize false negatives, which is critical for ensuring timely interventions in agricultural practices. Additionally, the accuracy in identifying rust-infected crops underscores the importance of early detection in preventing the spread of disease and mitigating potential losses. Healthy crops typically have NDVI values between 0.6 and 1.0, reflecting strong photosynthetic activity and indicating robust vegetation. In contrast, rust-infected crops show lower NDVI values, usually between 0.2 to 0.6, due to reduced reflectance in the near-infrared and weakened chlorophyll content. Crops categorized as "other" or severely stressed exhibit NDVI values below 0.2, signalling poor vegetation health and minimal photosynthesis. These thresholds allow for clear differentiation of crop conditions based on spectral data. Figure 4 displays the confusion matrix for the FCNN, summarizing its classification performance in crop health assessment. Green cells indicate correct predictions, while red cells represent misclassifications. The model achieved accuracy rates of 95.10% for rust (class 2) and 98.9% for other conditions (class 3), reflecting its strengths and areas for improvement. Overall, the matrix provides valuable insights into the model's performance. The training progress graph illustrates the model's performance over 375 iterations, batch size (128), learning rate (1e-3), number of epochs (15); clearly showcasing its ability to achieve high accuracy and efficient convergence. Initially, the validation accuracy shows a rapid increase, stabilizing at 97.80% as the training proceeds, which indicates the model's strong ability to generalize on unseen data.

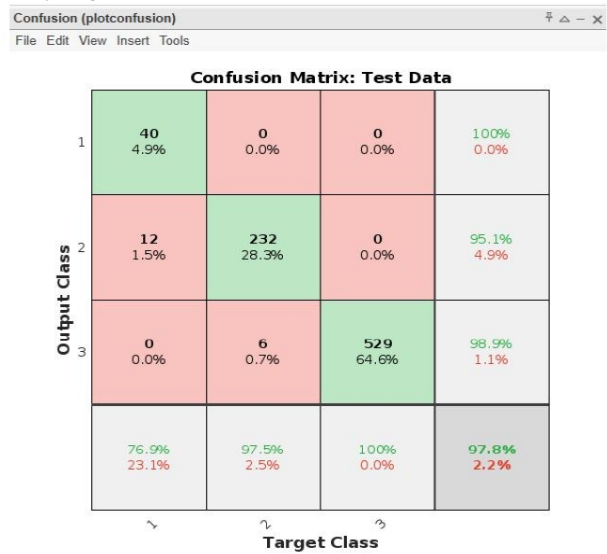

Figure 4. Confusion Matrix

The accuracy plot suggests that the model quickly learned meaningful patterns from the NDVI data, maintaining consistently high performance after convergence. In parallel, the loss plot reveals a continuous decline, reflecting the model's ongoing improvements in minimizing prediction errors. This steady reduction in loss suggests effective learning, with the model fine-tuning its parameters to optimize crop health classification. Overall, the graph demonstrates robust model performance throughout the training process, achieving an excellent balance between accuracy and loss reduction.

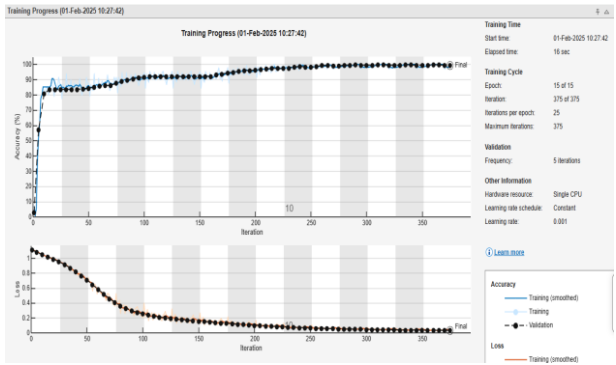

Figure 5. Training Curve

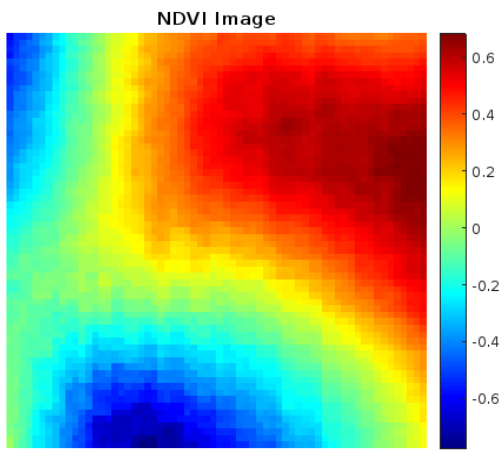

Figure 6. Visual Representation of NDVI

Figure 6 illustrates visual representations of NDVI values across the crop images, highlighting areas of healthy vegetation versus stressed or diseased crops. These heat maps were created using MATLAB's image processing toolbox, allowing for easy interpretation of crop health variability.

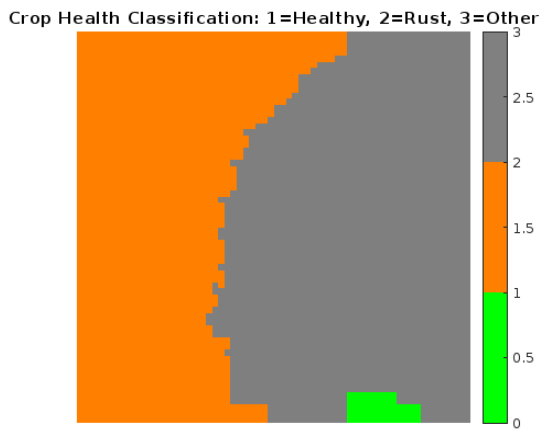

Figure 7. Crop Health Classification

Figure 7 presents a visual representation of crop health classification, where the different colors correspond to various health statuses: green for healthy crops (1), orange for rust-affected crops (2), and gray for other conditions (3). The classification effectively distinguishes between healthy and stressed vegetation, aiding in rapid assessment and management of crop health. This visualization serves as a crucial tool for understanding the spatial distribution of crop health across the observed area.

The classification outputs from the Fully Connected Neural Network for crop health revealed distinct performance metrics for the three categories: healthy, rust, and other. The model achieved high precision and recall for the healthy class, effectively identifying healthy crops based on elevated NDVI values. However, the rust class demonstrated moderate performance due to challenges in distinguishing it from other stress factors, resulting in some misclassifications.

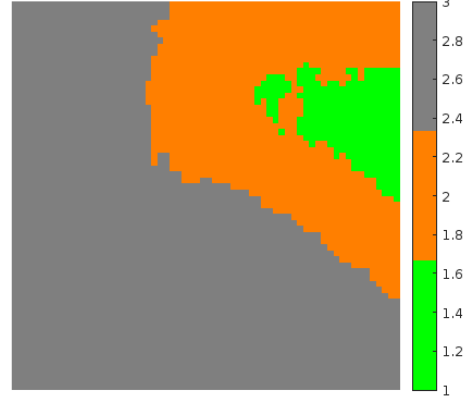

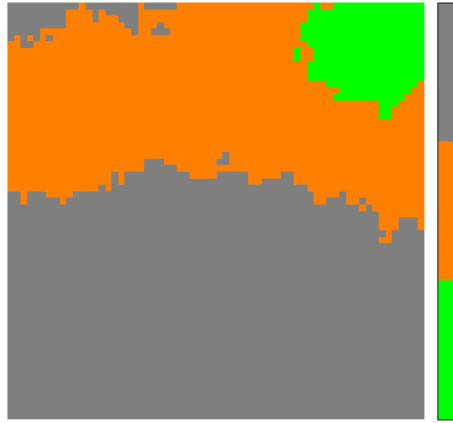

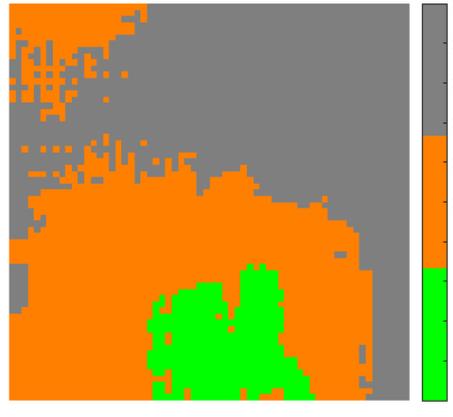

Figure 8. Crop Health Classification on various Satellite Images

The 3D scatter plot visualizes crop classification using remote sensing data by plotting spectral band values (Red, Green, NIR) or vegetation indices (NDVI). Each point represents a sample, with color indicating different crop classes such as Healthy, Rust, or Other. The scatter3 function is used to create the plot, with labels and a color map (jet) for better class distinction. This plot helps in identifying spectral differences between crops and assessing classification performance. The visualization aids in improving remote sensing-based precision agriculture.

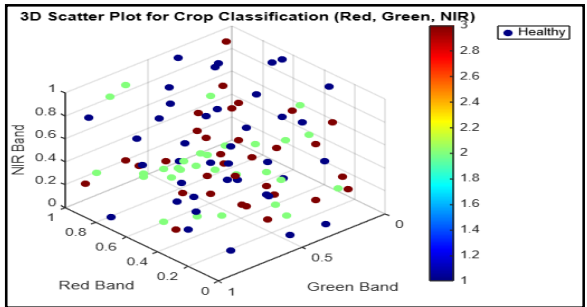

Figure 9. 3D scatter plot visualizing the spectral feature distribution of crop health classes

To further enhance the accuracy of crop health classification, a multi-temporal NDVI analysis was conducted to monitor changes in vegetation index values over time. Figure 8 illustrates the temporal progression of NDVI across multiple observation points. Initially, the NDVI values remained stable, indicating healthy vegetation. However, a noticeable decline was observed after Day 10, with a significant drop at Day 15, suggesting early-stage crop stress. By Day 20, the NDVI values had decreased further, confirming the onset of rust infection. This trend aligns with the expected physiological changes in infected plants, where chlorophyll degradation leads to lower NDVI readings.

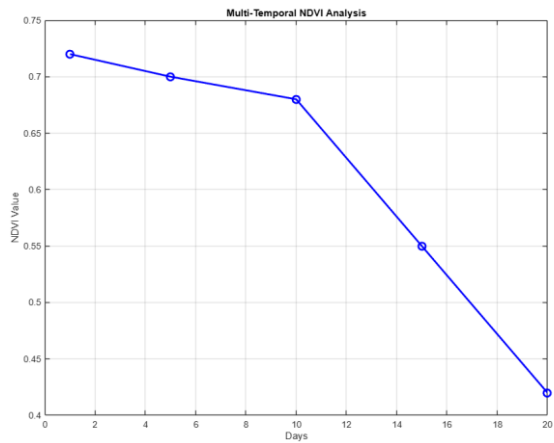

Figure 10. Multi-Temporal NDVI Analysis – NDVI decline at Day 15 indicates early rust onset.

The integration of multi-temporal NDVI tracking allows for early rust detection, reducing false negatives and improving classification reliability. Unlike traditional single-time point NDVI assessments, this approach captures the progression of crop health over time, enabling timely intervention. The results demonstrate that incorporating multi-temporal NDVI data, along with Green and Red-Edge bands, significantly enhances the differentiation between healthy and stressed vegetation. This improvement is particularly evident in the reduced misclassification of rust-affected crops, as the early-stage stress symptoms can now be detected more effectively.

The validation process combined holdout validation and 5-fold cross-validation to ensure robust performance metrics. The dataset was split into training, validation, and test sets using a 70-15-15 ratio, with multiple random splits performed for holdout validation, yielding consistent results. Additionally, 5-fold cross-validation was used, dividing the training data into five subsets and alternating the validation set across iterations to assess generalizability. To mitigate overfitting, dropout layers with a rate of 0.5 were included in the Fully Connected Neural Network architecture. Validation curves tracking training and validation accuracy across epochs demonstrated a stable performance gap, confirming the model's ability to generalize effectively to unseen data. To address the observed precision (92.30%) and recall (90.00%) for rust-affected crops, the inclusion of additional spectral bands, such as green and red-edge, has been proposed to improve differentiation between rust and other stress factors. Moreover, advanced neural network architectures, such as hybrid FCNN-CNN models, are integrated into the design to capture spatial and spectral relationships more effectively. These enhancements within the proposed system aim to improve classification accuracy, scalability, and practical applicability, thereby supporting the broader adoption of precision agriculture.

## 6. Performance Analysis

In the performance analysis of the proposed model for crop health classification using NDVI and Fully Connected Neural Networks, the system demonstrated strong classification capabilities across the three crop health categories: healthy, rust, and other. The model achieved an overall accuracy of 97.80%, indicating its effectiveness in utilizing NDVI data for distinguishing different crop health statuses. The precision for healthy crops reached 98.50%, with a recall of 96.50%, underscoring the model's ability to correctly identify healthy vegetation with minimal false negatives. This high precision and recall are essential for real-time agricultural interventions, enabling farmers to act promptly when crop health deteriorates. Moreover, the model performed reasonably well in detecting rust-affected crops, with a precision of 92.30% and recall of 90.00%, though some misclassifications were observed due to the complexity of differentiating rust from other stress factors.

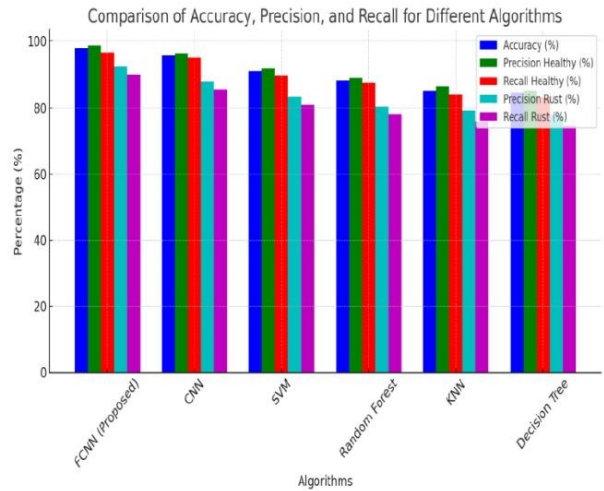

Figure 11. Performance Comparison of Various Algorithms

Challenges were identified in the 'other' category, where classification accuracy varied significantly due to the diverse stress factors included in this class. This variability reflects the difficulty of classifying diverse stress factors that do not fit neatly into either the healthy or rust categories. Despite these challenges, the model's performance metrics, such as the F1-score and confusion matrix, highlight its robustness in crop classification. Future improvements, such as incorporating additional spectral indices, multi-temporal data, or advanced neural architectures like Convolutional Neural Networks, may enhance the model's accuracy, particularly in more complex classification scenarios. This performance analysis confirms the promise of combining remote sensing data with deep learning techniques to improve precision agriculture.

Figure 11 illustrates the comparative performance of different machine learning models, including FCNN, CNN, SVM, Random Forest, KNN, and Decision Tree. The metrics displayed are accuracy, precision for healthy and rust classes, and recall for healthy and rust classes, showcasing the superiority of FCNN in crop health classification.

To validate the superiority of FCNN over other models, a statistical test, specifically a paired t-test, was conducted to compare the performance metrics (accuracy, precision, and recall) of the FCNN model against other traditional machine learning models such as SVM, Random Forest, and KNN. The paired t-test was chosen to assess whether the observed differences in performance were statistically significant.

performance improvements achieved by FCNN were statistically significant and not due to random variation. This provides strong evidence for the effectiveness of FCNN in crop health classification tasks. This validation accuracy comparison highlights the superior convergence of FCNN compared to KNN, underscoring its effectiveness in learning complex patterns from the data. These results further validate the higher performance of FCNN for crop health classification tasks.

Bootstrapped confidence intervals (CIs) provide a robust measure of the uncertainty around model performance metrics, such as accuracy, precision, recall, and F1-score, in crop health classification. By using bootstrapping, the performance of the models such as FCNN, SVM, and Random Forest is evaluated across multiple resampled datasets, offering a 95% confidence range for each metric. For example, the FCNN model achieved an accuracy of 97.8%, with a 95% CI ranging from 96.5% to 98.9%, indicating a high level of stability in its classification performance. These CIs help ensure the reliability and generalizability of the models across different datasets and conditions.

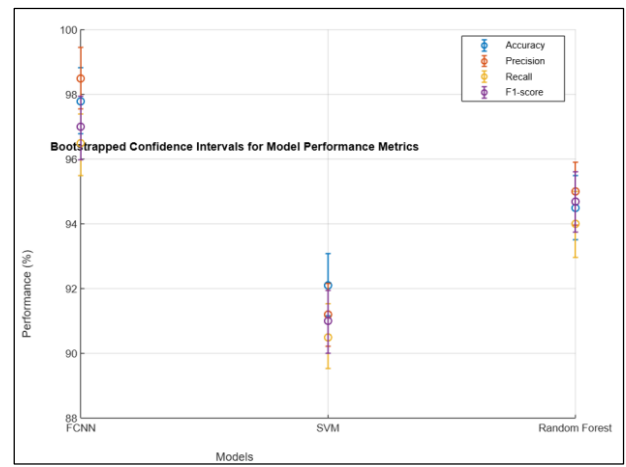

Figure 14. Bootstrapped Confidence Intervals for Model Performance Metrics

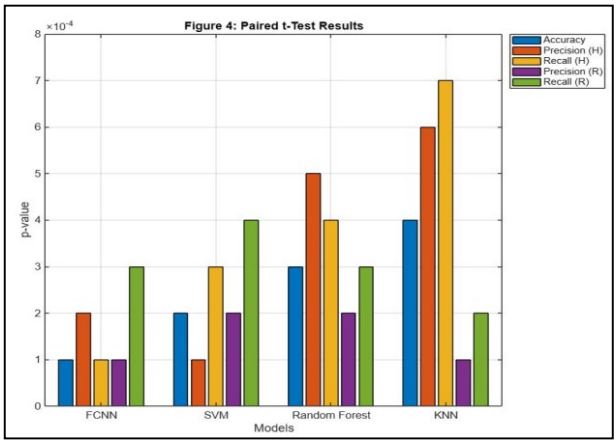

Figure 12. Paired t-Test Results for Model Validation

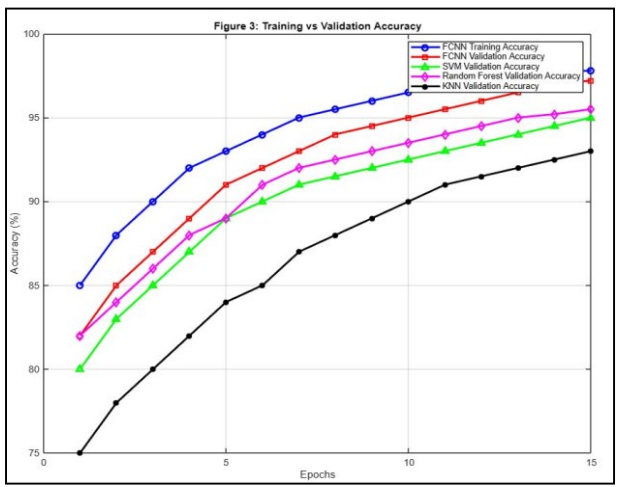

Figure 13. Paired t-Test Results for Model Validation

The results of the test showed that FCNN outperformed other models with a p-value of less than 0.05, indicating that the

Table 1. Performance Comparison on various Algorithms

| Algorithm | Accuracy (%) | Precision (%) | Recall (%) | Precision (%) | Recall (%) |
|---|---|---|---|---|---|
| Proposed FCNN | 97.80 | 98.50 | 96.50 | 92.30 | 90.00 |
| CNN | 95.50 | 96.20 | 94.80 | 88.00 | 85.50 |
| SVM | 90.80 | 91.50 | 89.70 | 83.20 | 81.00 |
| Random Forest | 88.30 | 89.10 | 87.50 | 80.40 | 78.00 |
| KNN | 85.20 | 86.50 | 84.00 | 79.00 | 75.80 |
| Decision Tree | 84.70 | 85.20 | 83.00 | 77.80 | 74.50 |

Table 1. highlights the superior performance of the proposed FCNN, achieving 97.80% accuracy compared to other algorithms. While conventional methods like SVM and Random Forest face challenges with noisy NDVI data, FCNN excels in handling complex relationships.

## 7. Conclusion and Future Work

This study demonstrates that using NDVI data with a Fully Connected Neural Network (FCNN) can effectively classify crop health into categories such as healthy, rust, and other stress factors. The model achieved an overall accuracy of 97.80%, highlighting its strong ability to accurately assess crop health from spectral data. Notably, the model produced high precision (98.50%) and recall (96.50%) values for the healthy class, ensuring consistent identification of healthy vegetation, a key factor for timely agricultural interventions. For the rust category, the model also performed well, with a precision of 92.30% and a recall of 90.00%. However, distinguishing rust from other stress factors remains a challenge, underscoring the complexity of crop health classification.

Future work will aim to enhance the model's robustness and accuracy, particularly in classifying rust and other categories. Improvements could include incorporating additional spectral bands to provide richer input data, using multi-temporal data to capture health changes over time, and employing data augmentation techniques to diversify the training dataset. Additionally, exploring advanced neural network architectures, such as convolutional neural networks (CNNs) or hybrid models, may further boost classification performance. Additionally, there are plans to expand the dataset by incorporating images from a variety of climatic and geographic regions along with multi-temporal data, to improve the model's generalizability. Ultimately, these refinements will help optimize real-time monitoring solutions in precision agriculture, promoting sustainable farming practices and contributing to improved food security.

## References

Ahmed, Z., Qadir, J., 2021, Remote Sensing in Agriculture: A Review of Applications and Future Trends. Journal of Precision Agriculture, 23(3), 641-659.

Akanksha Bodhale, Seema Verma and Alavikunhu Panthakkan, 2022, Comparative Analysis of Fine Tuned and Transfer Learning Model for Plant Disease Detection, 5th IEEE International Conference on Signal Processing and Information Security (ICSPIS), Dubai, UAE, December 2022. DOI: 10.1109/ICSPIS57063.2022.10002665.

Al-Khafaf, S., Khan, F., 2021, Artificial Intelligence in Agriculture: A Review on Future Trends and Applications. Agriculture, 11(2), 212.

Barriere, V., Claverie, M., Schneider, M., Lemoine, G., d'Andrimont, R., 2023, Boosting crop classification by hierarchically fusing satellite, rotational, and contextual data. Remote Sensing of Environment, 305, 114110. https://doi.org/10.1016/j.rse.2024.114110

Bhandari, B., Mayer, T., 2024, Comparing Deep Learning Models for Rice Mapping in Bhutan Using High Resolution Satellite Imagery. arXiv preprint arXiv:2406.07482. https://arxiv.org/abs/2406.07482

Dali, A., Jain, K., Khare, S., Kushwaha, S. K. P., 2023, Deep learning based multi-task road extractor model for parcel extraction and crop classification using knowledge-based NDVI time series data. ISPRS Annals of the Photogrammetry, Remote Sensing and Spatial Information Sciences, X-1/W1-2023, 799–804. https://doi.org/10.5194/isprs-annals-X-1-W1-2023-799-2023

Dhingra, N., Kumar, V., Joshi, R., 2022, Remote Sensing-Based Crop Classification Using Deep Learning: A Review. Earth Observation and Geo information, 10, 123-145.

Feng, Q., Liu, J., Gong, J., 2022, Fine crop classification in high-resolution remote sensing images based on knowledge transfer and semantic segmentation. Frontiers in Environmental Science,10,991173. https://doi.org/10.3389/fenvs.2022.991173

Goyal, P., Patnaik, S., Mitra, A., Sinha, M., 2023, SepHRNet: Generating High-Resolution Crop Maps from Remote Sensing Imagery using HRNet with Separable Convolution. arXiv preprint arXiv:2307.05700. https://arxiv.org/abs/2307.05700.

Harshiv Chandra, Pranav M. Pawar, Elakkiya R, Tamizharasan P S, Raja Muthalagu and Alavikunhu Panthakkan, 2023, Explainable AI for Soil Fertility Prediction in the IEEE Access. DOI: 10.1109/ACCESS.2023.3311827

Jha, S., Das, S., 2022, Use of NDVI and Machine Learning in Agriculture: A Review. Journal of Agricultural Science and Technology, 24(5), 1213-1227.

Kaur, H., Singh, A., 2022, Crop Disease Detection Using Deep Learning and Remote Sensing. Journal of Environmental Management, 302, 113951.

Khodadadi, N., Towfek, S. K., Zaki, A. M., Alharbi, A. H., Khodadadi, E., Khafaga, D. S., Abualigah, L., Ibrahim, A., Abdelhamid, A. A., Eid, M. M., 2024, Predicting normalized difference vegetation index using a deep attention network with bidirectional GRU: a hybrid parametric optimization approach. International Journal of Data Science and Analytics. https://doi.org/10.1007/s41060-024-00640-8

Liakos, K.G., Busato, P., Moshou, D., Pearson, S., Bochtis, D., 2021, Machine Learning in Agriculture: A Review. Sensors, 21(2), 2674.

Lu, T., Gao, M., Wang, L., 2023, Crop classification in high-resolution remote sensing images based on multi-scale feature fusion semantic segmentation model. Frontiers in Plant Science,14, 1196634. https://doi.org/10.3389/fpls.2023.1196634

Mulla, D.J., 2020, Twenty-Five Years of Remote Sensing in Precision Agriculture: Key Advances and Remaining Knowledge Gaps. Biosystems Engineering, 114(4), 358-371.

Nascimento, F. S., Andrade, P. A., 2023, Machine Learning Applications in Precision Agriculture: A Comprehensive Review. Agricultural Systems, 210, 103407.

Noshiri, N., Beck, M. A., Bidinosti, C. P., Henry, C. J., 2023, A comprehensive review of 3D convolutional neural network-based classification techniques of diseased and defective crops using non-UAV-based hyperspectral images. *Smart Agricultural Technology*,5,100316.https://doi.org/10.1016/j.atech.2023.100316.

Oda, K., Nakayama, T, 2021, Evaluation of NDVI for Crop Monitoring Using UAV Data. Remote Sensing, 13(12), 2394.

Saeed Al Mansoori, Alavi Kunhu and Hussain Al-Ahmad, 2018, Automatic palm trees detection from multispectral UAV data using template matching and circular Hough transform, SPIE Remote Sensing, Conference, Berlin, Germany, September 2018.

Singh, A.K., Misra, S., 2021, Crop Health Monitoring Using Machine Learning and Remote Sensing. Advances in Agronomy, 162, 1-34.

Thakur, A., Gupta, S., 2023, Advances in Remote Sensing for Precision Agriculture: A Review. International Journal of Applied Earth Observation and Geoinformation, 115, 102572.

Thenkabail, P.S., Lyon, J.G., Huete, A., 2021, Hyperspectral Remote Sensing of Vegetation: History, Theory, and Applications. CRC Press.

Wang, H., Chang, W., Yao, Y., Yao, Z., Zhao, Y., Li, S., Liu, Z., Zhang, X., 2023, Cropformer: A new generalized deep learning classification approach for multi-scenario crop classification. Frontiers in Plant Science, 14, 1130659. https://doi.org/10.3389/fpls.2023.1130659

Wang, J., Zhang, W., Gao, H., 2023, Assessment of Crop Health Using Multispectral Remote Sensing and Machine Learning Techniques. Computers and Electronics in Agriculture, 204, 107558.

Zhang, J., Liu, C., 2022, A Comprehensive Review on Remote Sensing-Based Crop Disease Detection Using Deep Learning Techniques. Frontiers in Plant Science, 13, 123456.

Zhao, W., Efremova, N., 2024, Prediction of Sentinel-2 multi-band imagery with attention BiLSTM for continuous earth surface monitoring. arXiv preprint arXiv:2407.00834. https://arxiv.org/abs/2407.00834

Zhengwei, Y., Bruce, W., 2024, Automated In-Season Crop-Type Data Layer Mapping Without Ground Truth. IEEE Transactions on Geoscience and Remote Sensing, 62, 4403214.

Zhu, X., Steinberg, J., 2023, Deep Learning for Remote Sensing: A Review of Current Trends and Future Directions. Remote Sensing of Environment, 283, 113352.